\documentclass[10pt]{wlscirep}
\usepackage[utf8]{inputenc}
\usepackage[T1]{fontenc}
\usepackage{amsmath}
\usepackage{hyperref}
\usepackage{lineno}
\usepackage{float}
\usepackage{orcidlink}

\title{Temporal and spatial evolution of the distribution related to the number of COVID-19 pandemic}

\author[1,*,+]{Peng Liu\orcidlink{0000-0002-9115-2055}}
\author[2,*,+]{Yanyan Zheng\orcidlink{0000-0002-1092-072X}}
\affil[1]{School of Information, Xi'an University of Finance and Economics, Xi'an 710100, Shaanxi, P.R. China}
\affil[2]{School of Management, Xi'an Polytechnic University, Xi'an 710048, Shaanxi, P.R. China}
\affil[*]{corresponding authors: \href{mailto:pengliuhep@outlook.com}{pengliuhep@outlook.com}, \href{mailto:yanzheng@whu.edu.cn}{yanzheng@whu.edu.cn}}
\affil[+]{these authors contributed equally to this work}

\begin{abstract}
This work systematically conducts a data analysis based on the numbers of
both cumulative and daily confirmed COVID-19 cases and deaths in a time span through April 2020 to June 2022 for over 200 countries around the world.
Such research feature aims to reveal the temporal and spatial evolution of the country-level distribution observed in COVID-19 pandemic,
and obtains some interesting results as follows.
(1) The distributions of the numbers for cumulative confirmed cases and deaths obey power-law in early stages of COVID-19 and stretched exponential function in subsequent course.
(2) The distributions of the numbers for daily confirmed cases and deaths obey power-law in early and late stages of COVID-19 and stretched exponential function in middle stages.
The crossover region between power-law and stretched exponential behaviour seems to depend on the evolution of "infection" event and "death" event.
Such observation implies a kind of important symmetry related to the dynamics process of COVID-19 spreading.
(3) The distributions of the normalized numbers for each metric show a temporal scaling behaviour in 2-year period,
and are well described by stretched exponential function.
The observation of power-law and stretched exponential behaviour in such country-level distributions suggests underlying intrinsic dynamics of a virus spreading process in human interconnected society.
And thus it is important for understanding and mathematically modeling the COVID-19 pandemic.
\end{abstract}

\begin{document}

\flushbottom
\maketitle

\thispagestyle{empty}
%\linenumbers
\section{Introduction}
COVID-19 global pandemic{\cite{WHO-COVID-19, outbreak-1, outbreak-2}} is caused by the spread of the infectious novel coronavirus SARS-CoV-2 (Severe Acute Respiratory Syndrome Coronavirus 2{\cite{name-it}}) in human interconnected society.
It has overpoweringly astonished SARS{\cite{WHO-SARS}} and MERS{\cite{WHO-MERS}} in terms of the geographical scope affected, the amount of people infected, and the economic recession worldwide.
As of 3rd June 2022, 
according to the report of the World Health Organization (WHO){\cite{WHO-COVID-19}},
COVID-19 has made at least 528 million people infected and 6 million people died.
These numbers may be under-reported dramatically in accordance with the latest estimation published in the Lancet{\cite{Lancet-death}}.
The ongoing pandemic has been posing a tremendous threat to global public health,  people's safety, and world economy.

It is observed from WHO Coronavirus (COVID-19) Dashboard that{\cite{WHO-COVID-19}}, by 3rd June 2022,
the cumulative number of confirmed COVID-19 cases varies widely across countries (just a few cases in some countries and up to 83 million cases in USA),
and that of deaths similarly spans in a huge range (lower to a few deaths in some countries and upper to 998 thousand deaths in USA).
There are also similar huge differences between countries in terms of the numbers of both daily confirmed cases and deaths.
One might attribute such huge country-level variations to several obvious things:
the significant difference of geography and population size of individual countries,
the different country-level containment policies including travel restrictions and other control measures,
the discrepancy in terms of testing ability and reporting policies across countries, 
and the number of people vaccinated being dramatically different across countries, 
among others.
Indeed, several researches have employed mathematical modeling tools to understand
the temporal and spatial development of this epidemic by considering non-pharmaceutical interventions{\cite{model-1, model-2, model-3, model-4, model-5, model-6, model-7}}.

However, we argue here that such huge variations are mainly dominated by the dynamics of COVID-19 spreading process in the human interconnected complex physical network,
in which a large number of subsystems are nonlinearly coupled.

Such complex system tends to exhibit a spatial and temporal scaling characteristic related to power-law,
which is also known as Pareto distribution or Zipf’s law{\cite{Power-Law-1, Power-Law-2, Power-Law-3}}, and defined as
\begin{equation}
\label{power-law}
P(x) \propto x^{-\mu}
\end{equation}
where $P\left(x\right)$ refers to the probability density at a value of random variable $x \left(x>0\right)$,
and $\mu$ is a positive parameter called shape parameter or tail index (or Pareto index when power-law is used to describe the distribution of wealth).
Power-law appears in a wide range of complex systems ranging from natural system to human society
{\cite{weath-dis1, weath-dis2, rainfall, Power-Law-3, settlement, return-1,return-2, human-mobility-1, human-mobility-2, BB-1, BB-2, BB-3, power-law-others}}.
Such universal phenomenon tends to indicate an intrinsic essential process underlying complex systems,
and might be explained by self-organized criticality{\cite{Power-Law-3, criticality}}, turbulence{\cite{turbulence-1, turbulence-2}}, Yule process{\cite{BB-1}}, and fractals theory{\cite{fractals}}.

Although power-law seems to show universality in many phenomena,
scholars have argued that power-law is overused since some observations of power-law in many data can not stand up to rigorous scrutiny{\cite{Stretched-exp}}.
Alternatively, stretched exponential function is proposed to describe that data{\cite{Stretched-exp}}.
The stretched exponential function is defined as
\begin{equation}
\label{stretched}
P(x) \propto e^{-\alpha x^{\beta}}
\end{equation}
where $P\left(x\right)$ refers to the probability density at a value of random variable $x \left(x>0\right)$,
$\alpha$ and $\beta$ are two key parameters to characterize the stretched exponential function.
The stretched exponential function has been often used to describe many phenomena, such as relaxation in disordered systems and survival related processes.

Actually, scholar has claimed the observation of power-law in the numbers of both cumulative confirmed COVID-19 cases and deaths for countries worldwide
in early stages of COVID-19 global pandemic{\cite{COVID-19-Power-Law}}.
Here we extend the previous study
by considering a longer time span of 2 years and more metrics including both cumulative and daily confirmed cases and deaths.
This new study aims to explore the temporal and spatial evolution of the behaviour in distributions of 4 metrics mentioned above.
Results show different behaviour from the findings of previous study.
Especially, a new finding that stretched exponential function can well describe the COVID-19 data is observed.

\section{Data analysis and results}
In this paper, we analyze the dataset of COVID-19 sourced from Our World in Data{\cite{Our-World-in-Data}},
which relies on the Johns Hopkins University dashboard and dataset maintained by a team at its Center for Systems Science and Engineering (CSSE){\cite{CSSE-1}}.
The datasets' details of Our World in Data and of CSSE can be accessed through the official website of Our World in Data{\cite{Our-World-in-Data}} and the publication in the journal of the Lancet Infectious Diseases{\cite{CSSE-2}}, respectively.
The dataset of Our World in Data has been daily updating since early stages of COVID-19 pandemic, and contains 63 metrics for more than 200 countries worldwide.
This paper focuses on 4 metrics for all countries listed in this dataset, which includes the numbers of cumulative confirmed cases, cumulative confirmed deaths, daily confirmed cases, and daily confirmed deaths.

Here we first explore the evolution of the country-level probability distributions of
both cumulative confirmed COVID-19 cases and deaths for more than 200 countries worldwide,
in a time span over 2-year period from early stages of COVID-19 through June 1, 2022,
as shown in Fig. \ref{fig1} and Fig. \ref{fig2}, respectively.
From Fig. \ref{fig1} we see that stretched exponential function $e^{-\alpha x^{\beta}}$ is well in agreement with the data points of cumulative confirmed cases
except for some points slightly deviating from fitting curves.
The parameters $\alpha$ and $\beta$ characterizing the stretched exponential function extracted from fitting approach  over 2-year period are consistent with each other within a statistical significance of $3\sigma$.
Fig. \ref{fig2} represents the data of cumulative confirmed deaths,
and shows a power-law behaviour on April 1, 2020 and stretched exponential behaviour in subsequent course.
The  parameters $\alpha$ and $\beta$ for cumulative confirmed deaths are consistent with those for cumulative confirmed cases within a statistical significance of $3\sigma$.
The previous similar study{\cite{COVID-19-Power-Law}} states that the country-level distributions of cumulative confirmed cases and deaths obey power-law at least in early stages of COVID-19 spreading.
Considering our analysis and the previous study{\cite{COVID-19-Power-Law},
we make a conclusion that the country-level distributions for cumulative confirmed cases and deaths behave as power-law in early stages of COVID-19 and a stretched exponential behaviour in subsequent course.
A crossover region that power-law gradually transforms to stretched exponential behaviour must exist in the course of COVID-19 pandemic evolution.
We also conclude that the crossover region for cumulative confirmed deaths appears latter than that of cumulative confirmed cases.

Fig. \ref{fig3} and Fig. \ref{fig4} present the country-level probability distributions of
both daily confirmed cases and deaths for more than 200 countries worldwide,
in a time span over 2-year period from early stages of COVID-19 through June 1, 2022.
Similar behaviour as illustrated in Fig. \ref{fig1} and Fig. \ref{fig2} is observed for such daily metrics. 
However, distinctly, the distributions of daily confirmed deaths show power-law behaviour in early stages of COVID-19 and recent days, and stretched exponential behaviour in middle stages.
Comparing with cumulative confirmed deaths, the distribution of daily confirmed deaths shows a very slow tendency transforming from power-law to stretched exponential behaviour.
The distributions of daily confirmed cases before April 1, 2022 are checked and obey power-law in early stages of COVID-19 pandemic.

From Fig. \ref{fig3} and Fig. \ref{fig4}, for  daily metrics,
we could make an important conclusion that the country-level distribution obeys power-law in early stages of COVID-19 pandemic,
and then gradually transforms to stretched exponential function in middle stages,
and finally go back to power-law in late stages.
The crossover region between power-law and stretched exponential function in early and late stages for such pandemic seems to depend on the evolution of "infection" event and "death" event.
In early stages of pandemic, the "infection" event occurs before the "death" event.
Such fact may account for the observation that the crossover region for daily confirmed deaths is significantly later than that of daily confirmed cases.
In late stages of pandemic, the "death" event disappears before "infection" event.
This scenario may explain the observation that
the distribution for daily confirmed deaths on June 1, 2022 obeys power-law,
but the distribution for daily confirmed cases on same day still follows stretched exponential behaviour.
We could make a prediction here that the distribution for daily confirmed cases in very late stages of pandemic will go back to power-law.
Such observation implies a kind of important symmetry related to the dynamics of COVID-19 spreading process.

%%%%%%%%%%%%%%%%%%%%%%%%%%%%%%%%%%%%%%%%%%%%%%%%%%
\begin{figure}[H]
\centering
\includegraphics[width=0.83\linewidth]{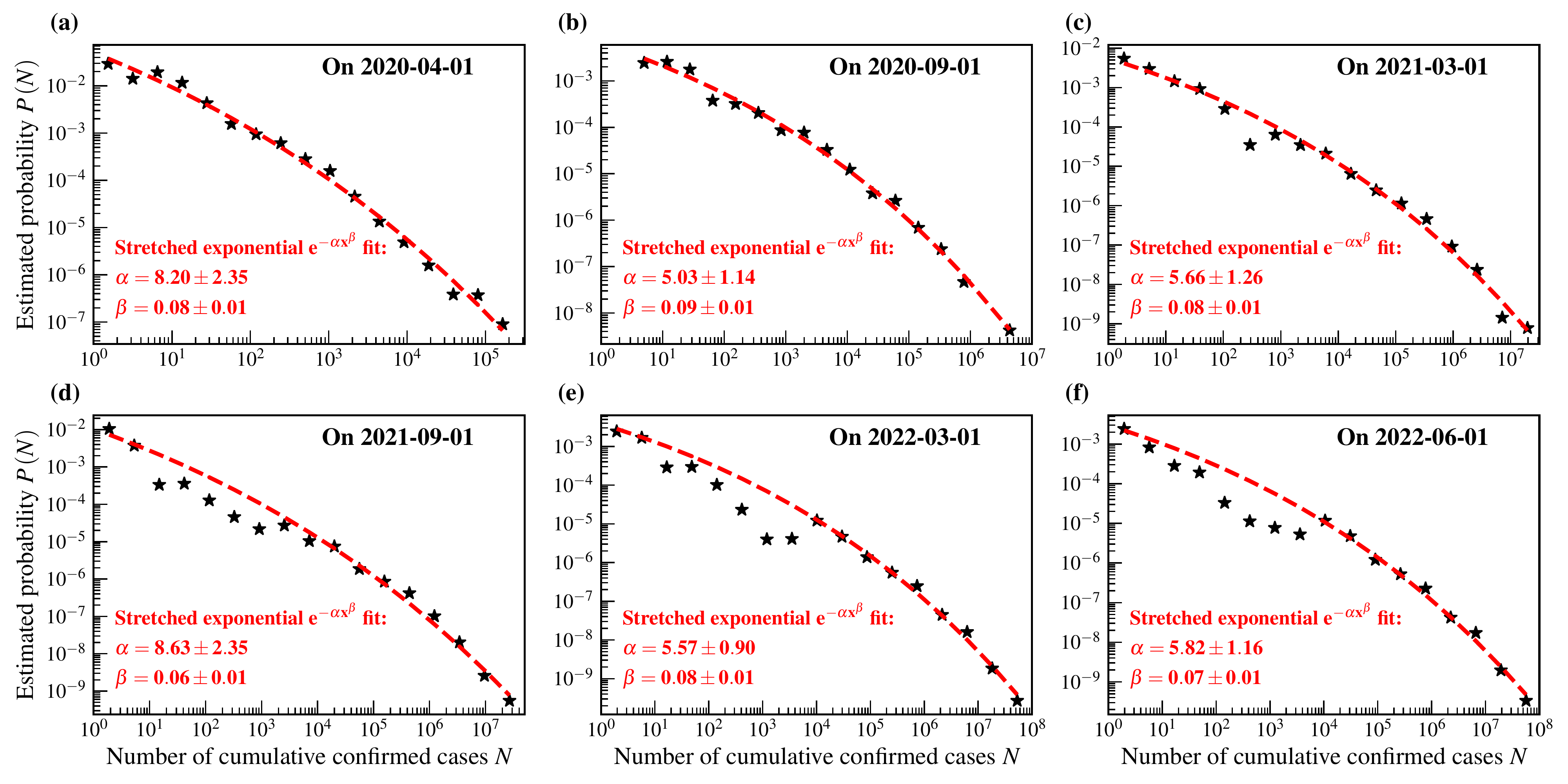}
\caption{
\textbf{Country-level distributions of the number of cumulative confirmed cases in log-log axes.}
The data points represent estimated probability $P\left(N\right)$ of a country to have N cumulative confirmed cases.
The red dashed curves stand for the results of fitting to data points using stretched exponential function $e^{-\alpha x^{\beta}}$.
Two key parameters $\alpha$ and $\beta$ with standard errors extracted from fitting are shown by red text.
The stretched exponential function is well describing the data in entire range of N except for some points deviating from fitting curves.
}
\label{fig1}
\end{figure}

\begin{figure}[H]
\centering
\includegraphics[width=0.83\linewidth]{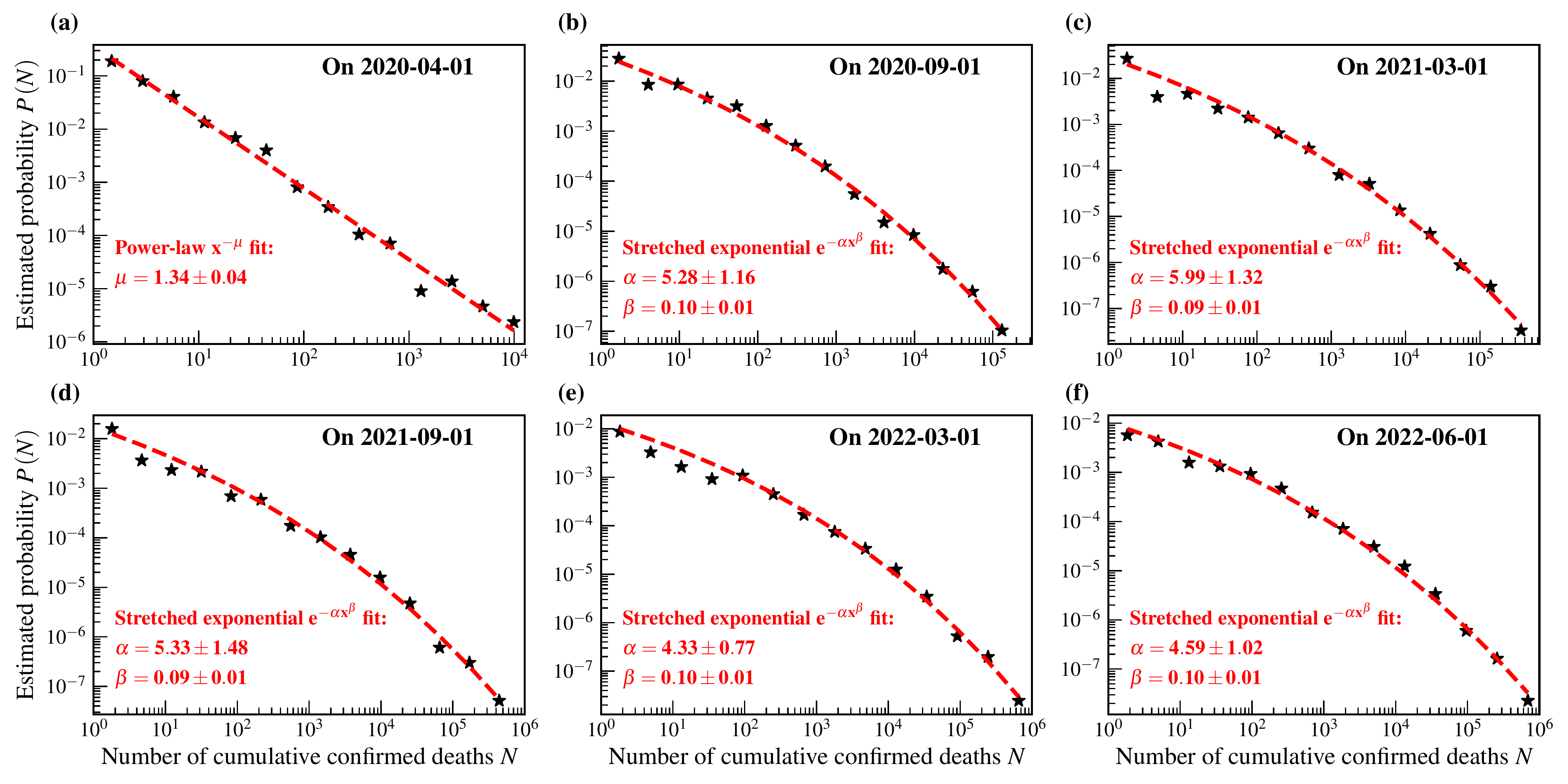}
\caption{
\textbf{Country-level distributions of the number of cumulative confirmed deaths in log-log axes.}
The data points represent estimated probability $P\left(N\right)$ of a country to have N cumulative confirmed deaths.
The red dashed straight line and curves stand for the results of fitting to data points using power-law $x^{-\mu}$ and stretched exponential function $e^{-\alpha x^{\beta}}$, respectively.
The fitting parameters with standard errors are shown by red text.
Obvious power-law behaviour during early days and stretched exponential behaviour during subsequent course, are observed here.
}
\label{fig2}
\end{figure}

\begin{figure}[H]
\centering
\includegraphics[width=0.83\linewidth]{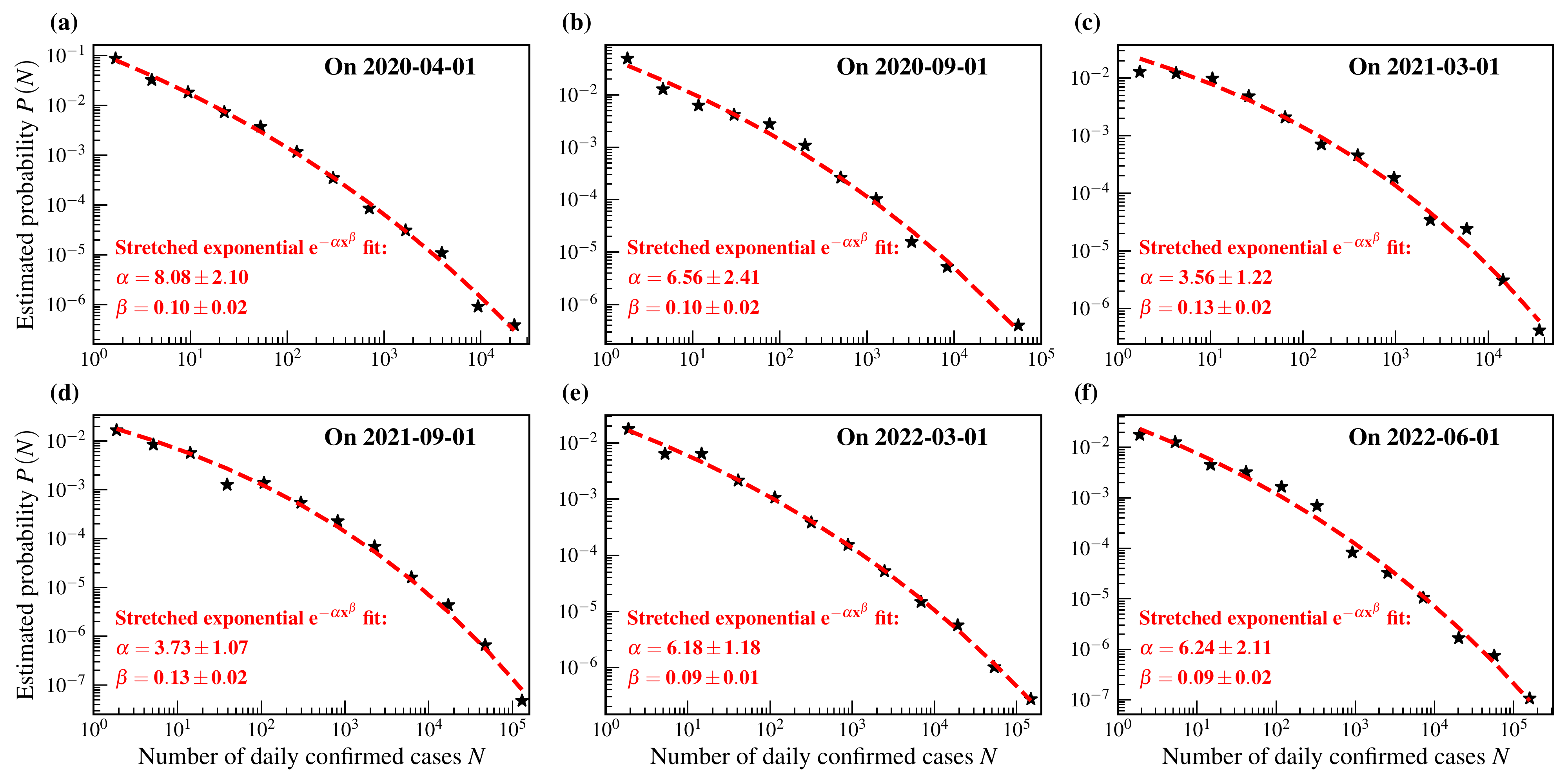}
\caption{
\textbf{Country-level distributions of the number of daily confirmed cases in log-log axes.}
The data points represent estimated probability $P\left(N\right)$ of a country to have N daily confirmed cases.
The red dashed curves stand for the results of fitting to data points using stretched exponential function $e^{-\alpha x^{\beta}}$.
Two key parameters $\alpha$ and $\beta$ with standard errors extracted from fitting are shown by red text.
The stretched exponential function is well describing the data in entire range of N.
}
\label{fig3}
\end{figure}

\begin{figure}[H]
\centering
\includegraphics[width=0.83\linewidth]{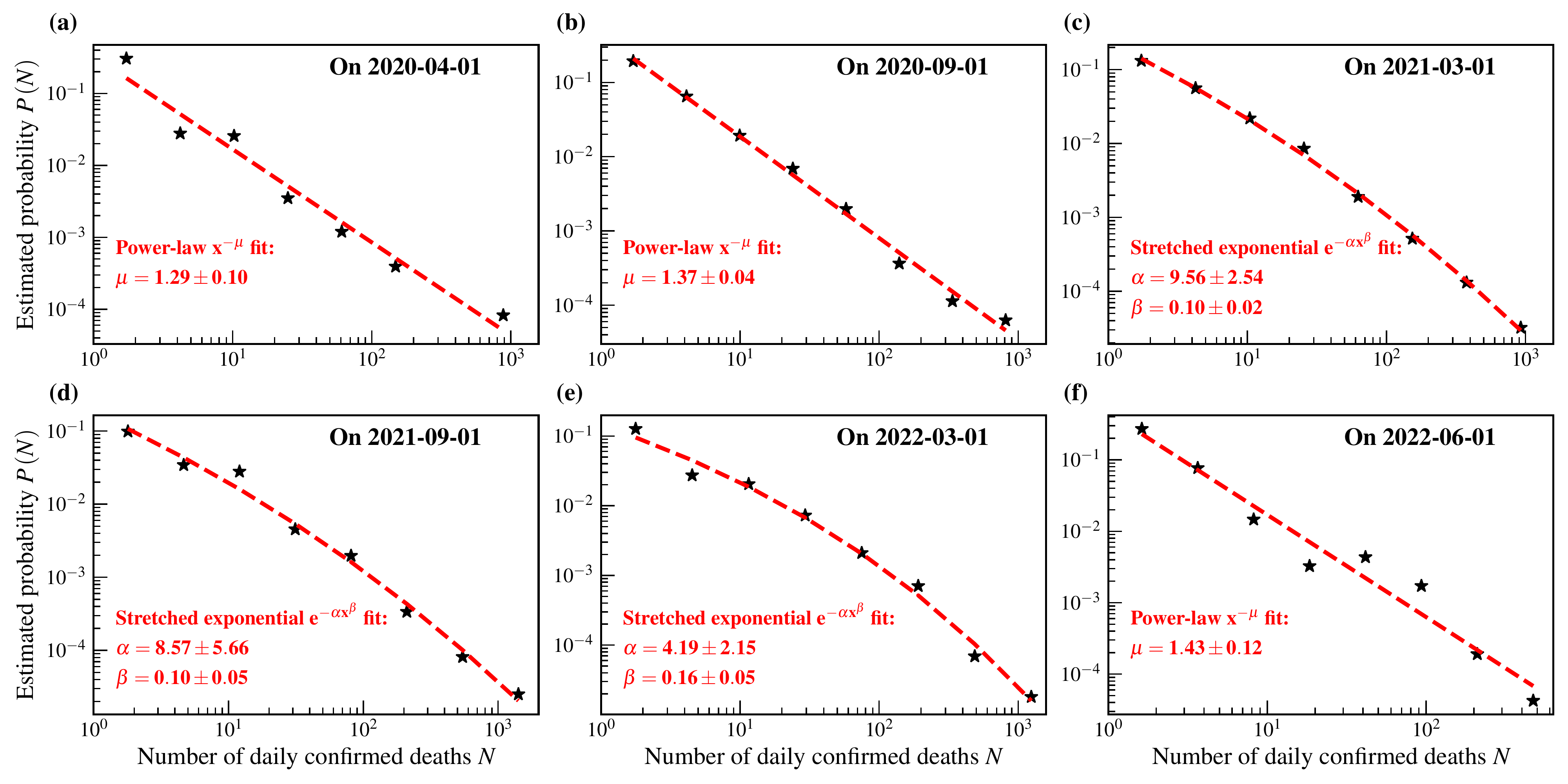}
\caption{
\textbf{Country-level distributions of the number of daily confirmed deaths in log-log axes.}
The data points represent estimated probability $P\left(N\right)$ of a country to have N daily confirmed deaths.
The red dashed straight lines and curves stand for the results of fitting to data points using power-law $x^{-\mu}$ and stretched exponential function $e^{-\alpha x^{\beta}}$, respectively.
The fitting parameters with standard errors are shown by red text.
Obvious power-law behaviour during early days and recent days and stretched exponential behaviour during middle stages, are observed here.
}
\label{fig4}
\end{figure}

To compare the country-level distributions on different dates,
here we explore the possibility of a universal scaling behaviour across dates using the normalized COVID-19 numbers as shown in Fig. \ref{fig5}.
The normalized number $n$ on a day is defined as
\begin{equation}
\label{normalized}
n(t) = \frac{N(t)}{\sqrt{E\left(N^2(t)\right)}}
\end{equation}
where $N(t)$ refers to a metric analyzed in this work on a specific date $t$,
and $\sqrt{E\left(N^2(t)\right)}$ is the variance of $N(t)$ when $E\left(N(t)\right) = 0$.
$E(\bullet)$ is the mean value of numbers across countries.
It is obvious that the probability density distributions of normalized numbers for different dates collapse together as shown in Fig. \ref{fig5}.
Thus, the normalized numbers across dates starting from 1st April 2020 to 1st June 2022 are combined day-by-day.
The distribution of combined numbers shows similar behaviour with the original data.
Stretched exponential function $e^{-\alpha x^{\beta}}$ is employed to fit the combined data points, as shown with red curves in Fig. \ref{fig5}.
From fitting results we see that the stretched exponential function is well describing the distribution of normalized numbers and a temporal scaling behaviour is obviously observed here.
%%%%%%%%%%%%%%%%%%%%%%%%%%%%%%%%%%%%%%%%%%%%%%%
\begin{figure}[H]
\centering
\includegraphics[width=0.83\linewidth]{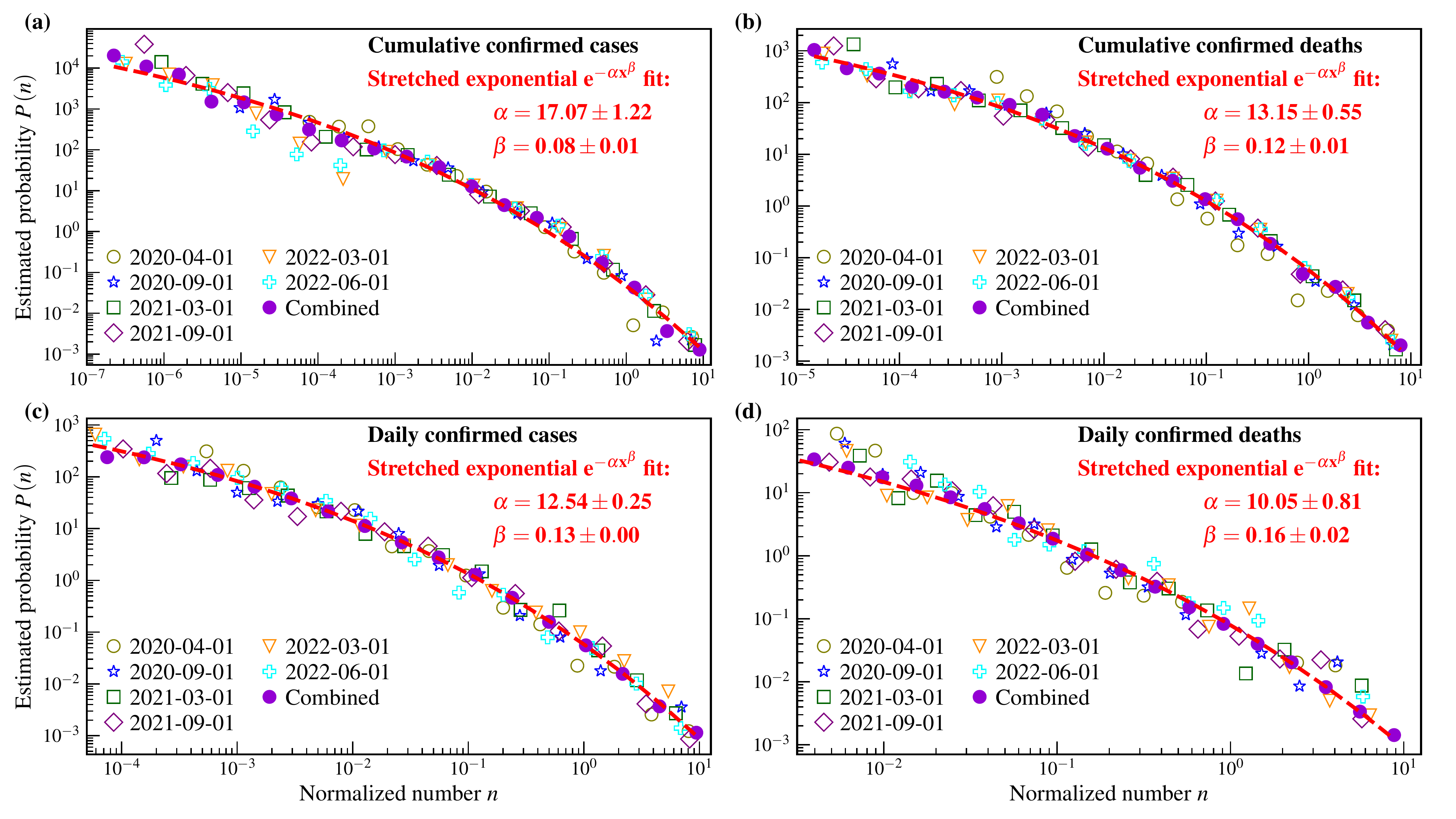}
\caption{
\textbf{Temporal scaling in the country-level distributions of the normalized numbers in log-log axes.}
The open markers with different colors represent  estimated probability $P\left(n\right)$ of a country to have normalized numbers $n$ on different dates.
The solid violet circles combines all data day-by-day starting from April 1, 2020 to June 1, 2022.
The red dashed curves are the results of fitting to combined data using stretched exponential function $e^{-\alpha x^{\beta}}$.
The parameters $\alpha$ and $\beta$ with standard errors extracted from fitting are shown by red text.
This figure demonstrates that the country-level distributions of normalized numbers on different dates collapse together,
and a temporal scaling behaviour related to stretched exponential function is obviously observed here.}
\label{fig5}
\end{figure}
 
\section{Conclusions}
The distribution obeying power-law or stretched exponential function appears in many complex systems ranging from natural system to human interconnected society.
This paper carefully analyzes the COVID-19 dataset sourced from Our World in Data
to explore the temporal and spatial evolution of the country-level distributions related to the system in which human virus spreads.
Specifically, the numbers of cumulative confirmed COVID-19 cases and deaths and the daily confirmed cases and deaths for over 200 countries around the world
over 2-year period through 1st April 2020 to 1st June 2022 are studied here.
The main findings of this work are as follows.
(1) The country-level probability distributions for cumulative confirmed cases and deaths show similar behaviour.
The distributions obey power-law in early stages of this pandemic and then change to stretched exponential behaviour in subsequent course.
However, the crossover region between power-law and stretched exponential behaviour for cumulative confirmed deaths appears later than that of cumulative confirmed cases.
(2) The country-level probability distributions for daily confirmed cases and deaths show similar behaviour.
Such distributions obey power-law in early and late stages of this pandemic, and behave as stretched exponential behaviour in middle stages.
The crossover region from power-law to stretched exponential behaviour (or from stretched exponential behaviour to power-law) seems to depend on the evolution of "infection" event and "death" event.
Such observation implies a kind of important symmetry related to the dynamics of COVID-19 spreading process.
(3) The country-level distributions of normalized COVID-19 numbers on different dates in a time span of 2-year period collapse together for the 4 metrics,
and thus a temporal scaling behaviour is observed in at least 2-year period.
The distributions of combined normalized numbers are well described by a stretched exponential function.

The temporal and spatial evolution of power-law and stretched exponential function observed in this paper indicates intrinsic dynamics of underlying COVID-19 spreading process in human interconnected society.
Thus, this work is important for understanding and mathematically modeling the dynamics of COVID-19 pandemic{\cite{COVID-19-Power-Law}},
especially under the circumstance that some countries are stopping tracking COVID-19{\cite{STOPING}}.

\section*{Acknowledgements}
This work was supported partially by the Education Department of Shaanxi Provincial Government (Contract No. 19JK0355)
and the Shaanxi Science and Technology Department (Contract No. 2020KRM001).
The authors would like to thank the anonymous reviewers for their insightful comments which make the quality of this paper better.

\section*{Author contributions}
All authors made important contributions to this publication in acquisition of data and data analysis.
Peng Liu wrote the manuscript.
All authors reviewed and approved the submitted manuscript.

\section*{Competing interests}
The authors declare no competing interests.

\end{document}